\documentclass{aa}
\usepackage{graphicx}
\begin{document}
%

   \title{Discovery of optically faint obscured quasars with Virtual Observatory tools}


   \author{Paolo Padovani
          \inst{1}
          \and
          Mark G. Allen\inst{2}
          \and
          Piero Rosati\inst{3}
          \and
          Nicholas A. Walton\inst{4}
          }

   \offprints{Paolo Padovani}

   \institute{ST-ECF, European Southern Observatory, Karl-Schwarzschild-Str. 2,
             D-85748 Garching bei M\"unchen, Germany\\
              \email{Paolo.Padovani@eso.org}
         \and
             Centre de Donn\'es astronomiques de Strasbourg (UMR 7550), 11 
             rue de l'Universit\'e, F-67000 Strasbourg, France\\ 
             \email{allen@astro.u-strasbg.fr}
         \and
             European Southern Observatory, Karl-Schwarzschild-Str. 2,
             D-85748 Garching bei M\"unchen, Germany\\
              \email{Piero.Rosati@eso.org}
         \and
             Institute of Astronomy, Madingley Road, Cambridge CB3 0HA, UK\\
             \email{naw@ast.cam.ac.uk}
             }

   \date{Received ; accepted }

\abstract{We use Virtual Observatory (VO) tools to identify optically faint,
obscured (i.e., type 2) active galactic nuclei (AGN) in the two Great
Observatories Origins Deep Survey (GOODS) fields. By employing publicly
available X-ray and optical data and catalogues we discover 68 type 2 AGN
candidates. The X-ray powers of these sources are estimated by using a
previously known correlation between X-ray luminosity and X-ray-to-optical
flux ratio. Thirty-one of our candidates have high estimated powers 
($L_{\rm x} >
10^{44}$ erg/s) and therefore qualify as optically obscured quasars, the
so-called ``QSO 2''.  Based on the derived X-ray powers, our candidates are
likely to be at relatively high redshifts, $z\sim 3$, with the QSO 2 at
$z\sim4$. By going $\sim3$ magnitudes fainter than previously known type 2 AGN
in the two GOODS fields we are sampling a region of redshift -- power space
which was previously unreachable with classical methods. Our method brings to
40 the number of QSO 2 in the GOODS fields, an improvement of a factor
$\sim 4$ when compared to the only 9 such sources previously known. We derive
a QSO 2 surface density down to $10^{-15}$ erg cm$^{-2}$ s$^{-1}$ in the 
$0.5 - 8$ keV band of $\ga 330$
deg$^{-2}$, $\sim 30\%$ of which is made up of previously known sources. This
is larger than current estimates and some predictions and suggests that the surface
density of QSO 2 at faint flux limits has been underestimated. This work
demonstrates that VO tools are mature enough to produce cutting-edge science
results by exploiting astronomical data beyond ``classical'' identification
limits ($R \la 25$) with interoperable tools for statistical identification of
sources using multiwavelength information.
   
}
   \maketitle
%

\section{Introduction}

The unified model for active galactic nuclei (AGN) is largely accepted
(e.g., Urry \& Padovani \cite{UP95}; see also the very recent results by
Jaffe et al. \cite{Jaffe04}). The apparent disparate properties and
nomenclature of active galaxies can be explained by the physics of black
hole, accretion disk, jet, and obscuring torus convolved with the geometry
of the viewing angle. Type 1 sources are those in which we have an 
unimpeded view
of the central regions and therefore exhibit the straight physics of AGN with
no absorption. Type 2 objects arise when the view is obscured by the
torus. While many examples of local, and therefore relatively low-power, type
2 AGN are known (the Seyfert 2s), it has been debated if their high-power
counterparts, that is optically obscured, radio-quiet type 2 QSO, exist. 
Indeed, until
very recently, very few, if any, examples of this class were known. Apart from
their importance for AGN models, type 2 sources are expected to make a
significant fraction of the X-ray background (see, e.g., Comastri et
al. \cite{Comastri01}) and are therefore also cosmologically very
relevant. These sources are heavily reddened and therefore fall through the
``standard'' (optical) methods of quasar selection. The hard X-rays, however,
are thought to be able to penetrate the torus. Type 2 QSO, therefore, should
have narrow, if any, permitted lines (and might look like normal galaxies in
the optical/UV band), powerful hard X-ray emission, and, in some cases, a high
equivalent width Fe K line (e.g., Norman et al. \cite{Norman02}).

In this paper we use Virtual Observatory (VO) tools to identify 68 type 2 AGN
candidates in the two Great Observatories Origins Deep Survey (GOODS) fields
(Giavalisco et al. \cite{Giava04}), $\sim 1/2$ of which qualify as QSO 2
candidates. Based on the properties of already known sources, we expect the
large majority of these to be obscured quasars whose identification is only
possible through their X-ray emission.

VO initiatives are now at a stage where prototype tools can be utilised to
produce scientific results.  Real gains have been made in the areas of
accessing and describing remote data sets, manipulating image and catalogue
data, and performing remote calculations in a fashion similar to grid
computing. These prototype tools are enabled by the VO infrastructure and
interoperability standards that are being developed cooperatively by all the
VO projects under the auspices of the IVOA\footnote{\tt http://www.ivoa.net}.
VO software is expected to mature significantly over the next 1-2 years as the
VO projects progress from demonstrations to building robust systems. In this
paper we have taken advantage of the first interoperability gains to produce
the ``first science'' for the VO.

In Section 2 we present the data we have used, Section 3 discusses the tools
we employed, while Section 4 describes our method. Section 5 presents our
results, which are discussed in Section 6, while Section 7 summarizes our
conclusions. Throughout this paper spectral indices are written $S_{\nu}
\propto \nu^{-\alpha}$ and we adopt a cosmological model with $H_0 = 70$ km
s$^{-1}$ Mpc$^{-1}$, $\Omega_{\rm M} = 0.3$, and $\Omega_{\rm \Lambda} = 0.7$.


\section{The data}

We use for our purposes the two GOODS fields (Giavalisco et al.
\cite{Giava04}), namely the Hubble Deep Field-North (HDF-N) and the Chandra
Deep Field-South (CDF-S). Since GOODS includes some of the deepest
observations from space- and ground-based facilities, these are the most
data-rich, deep survey areas on the sky. The GOODS field centres (J2000.0) are
12$^h$36$^m$55$^s$, +62$^{\circ}$14$^{\prime}$15$^{\prime\prime}$ for the
HDF-N and 3$^h$32$^m$30$^s$, $-27^{\circ}$48$^{\prime}$20$^{\prime\prime}$ for
the CDF-S. Each field provides an area of approximately 10$^{\prime}$ $\times$
16$^{\prime}$.

Deep X-ray (Chandra) catalogues are available for a larger region around both
fields (Alexander et al. \cite{Alexander03}; Giacconi et al.
\cite{Giacconi02}). For consistency, we use the catalogues produced by
Alexander et al. (\cite{Alexander03}) for both the 2 Ms HDF-N and 1 Ms CDF-S
data. These include 503 (HDF-N) and 326 (CDF-S) objects respectively, for a
total of 829 sources. The data cover the $0.5 - 8.0$ keV band and the
catalogues provide counts and fluxes in various sub-bands. Note that, due to
the twice as long exposure time, the HDF-N reaches fainter X-ray 
fluxes and includes a ($54\%$) larger number of sources.

In the optical we use the publicly available GOODS Hubble Space Telescope
(HST) Advanced Camera for Surveys (ACS) data (proposal ID 9\,583). The
observations consist of imaging in the F435W, F606W, F775W, and F850LP
passbands, hereafter referred to as $B, V, i,$ and $z$, respectively. We use
here version v1.0 of the reduced, calibrated, stacked, and mosaiced images and
catalogues as made available by the GOODS team\footnote{\tt
http://www.stsci.edu/science/goods/}.

Finally, identifications and redshifts through optical spectroscopy are
available from Barger et al. (\cite{Barger03}) and Szokoly et al.
(\cite{Szokoly04}) for the HDF-N and CDF-S respectively. We note that $\sim
56\%$ of the X-ray sources in the GOODS fields have spectroscopic redshift
determinations. These sources are necessarily relatively bright, with $\langle
I \rangle \sim 21$ (HDF-N; only $3\%$ fainter than 24th mag) and $\langle R
\rangle \sim 22$ (CDF-S; only $5\%$ fainter than 25th mag).

\section{Virtual Observatory Tools}

Astronomy is at a turning point. Major breakthroughs in telescope,
detector, and computer technology allow astronomical surveys to produce
massive amounts of images, spectra, and catalogues. These datasets cover the
sky at all wavelengths from $\gamma$- and X-rays, optical, infrared,
through to radio. The VO is an international,
community-based initiative, to allow global electronic access to available
astronomical data, both space- and ground-based. The VO aims also to enable
data analysis techniques through a coordinating entity that will provide
common standards, wide-network bandwidth, and state-of-the-art analysis
tools.

The Astrophysical Virtual Observatory Project (AVO)\footnote{\tt
http://www.euro-vo.org/} is conducting a research and demonstration programme
on the scientific requirements and technologies necessary to build a VO for
European astronomy. The AVO has been jointly funded by the European Commission
(under FP5 - Fifth Framework Programme) with six European organisations
participating in a three year Phase-A work programme.

The AVO project is driven by its strategy of regular scientific demonstrations
of VO technology. For this purpose an ``AVO prototype'' has been built. The
prototype consists of a suite of interoperable software, plus a set of
conventions or standards for accessing remote data, and for launching remote
calculations. The main component of the software is based on the CDS Aladin
visualisation interface (Bonnarel et al. \cite{Bonnarel2000}). This prototype
VO portal (v. 1.003-$\beta$) allows efficient interactive manipulation of
image and catalogue data, and provides access to remote data archives and
image servers via the GLU registry of services\footnote{\tt
http://simbad.u-strasbg.fr/glu/glu.htx}.

The Aladin image server is an example of such a VO service. It describes the
images stored in the Aladin database using a data model (Images Distribu\'ees
H\'et\'erog\`enes pour l'Astronomie; IDHA\footnote{\tt
http://cdsweb.u-strasbg.fr/idha.html}), and provides image cutouts on
request. In this paper we make heavy use of cutouts of the GOODS data
available via this service. The prototype is also interoperable with the other
long standing CDS Vizier and SIMBAD services (Ochsenbein, Bauer, \& Marcout
\cite{Ochsenbein2000}), and significant interoperability gains are achieved by
use of the VOTable\footnote{\tt http://cdsweb.u-strasbg.fr/doc/VOTable} format
for astronomical tables.

The prototype includes a catalogue cross matching service. This service allows
positional cross matching of two catalogues to find the best matched source,
all matching sources, or sources not matching within a given threshold radius
(in arcseconds). These three modes, plus the ability to compare directly with
the images from which the catalogues were generated, make for an extremely
efficient tool for which to perform cross matches and check for multiple, or
aberrant matches.

In addition to the AVO software, we have also made intensive use of the
Starlink topcat tool\footnote{\tt http://www.starlink.ac.uk/topcat/}, and the
VOIndia VOPlot plugin\footnote{\tt http://vo.iucaa.ernet.in/$\sim$voi/voplot.htm}
to the prototype.

\section{The method}

The two key physical properties that we use to identify type 2 AGN 
candidates are that they be obscured, and that they have sufficiently
high power to be classed as an AGN and not a starburst. To find these
candidates, we use a relatively simple method based on the X-ray and
optical fluxes.

Optical data alone are not sufficient for this purpose because at the
redshifts that we expect to find these sources the nuclear, AGN emission can
be diluted by the host galaxy.  Indeed, Moran, Filippenko \& Chornock
(\cite{Moran02}) have shown that $\sim 60\%$ of local Seyfert 2 galaxies
would be classified as normal galaxies if no decomposition of their optical
emission were available. However, AGN reveal themselves by their hard X-ray
emission and power. Ideally, to find type 2 AGN, one would need X-ray spectra
to select sources with flat spectral indices, indicative of absorption. The
typical count rates of the X-ray sources in the GOODS fields are however
relatively low for detailed spectral fitting and therefore we select sources
based on their X-ray hardness ratio, a measure of the fraction of hard photons
relative to soft photons.

To estimate the X-ray power for candidate type 2 sources we use the
correlation described by Fiore et al. (\cite{Fiore03}) between the $f(2 - 10
keV)/f(R)$ ratio and the hard X-ray power (see their Fig. 5). The basis
of this correlation is the fact that the $f(2 - 10 keV)/f(R)$ ratio is
roughly equivalent to the ratio between the nuclear X-ray power and the
host galaxy R band luminosity. Since the host galaxy R band luminosities 
(unlike the X-ray power) show only a modest amount of scatter, this flux ratio
is a good indicator of X-ray power. 

Each of the steps in our method, including our technique for
identifying the optical counterparts of the X-ray sources, are described 
in detail below.

\subsection{Selecting absorbed sources}\label{abs}

\begin{figure}
   \centering
   \resizebox{\hsize}{!}{\includegraphics{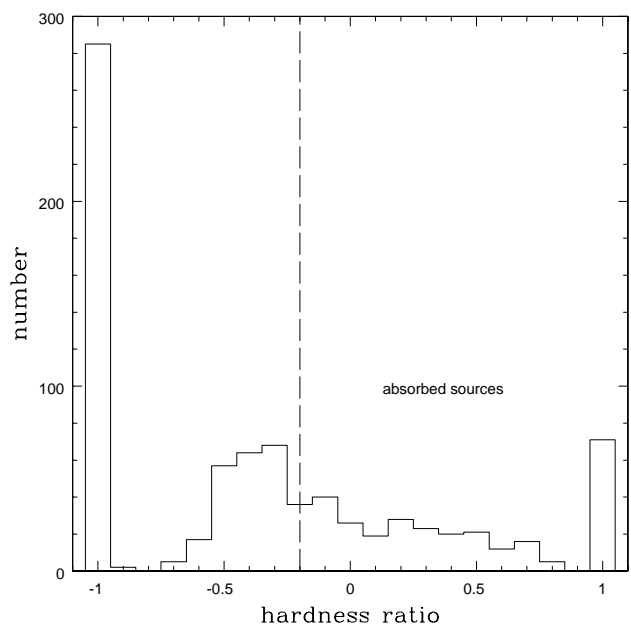}}
      \caption{The distribution of hardness ratios in the Alexander et al.
               (\cite{Alexander03}) catalogues. We define as absorbed sources 
               to the right of the dashed line, that is having hardness ratio
               $HR \ge -0.2$.}
         \label{HR}
   \end{figure}

The Alexander et al. (\cite{Alexander03}) catalogues provide counts in various
X-ray bands. We define the hardness ratio $HR = (H-S)/(H+S)$, where $H$ is the
hard X-ray counts ($2.0 - 8.0$ keV) and $S$ is the soft X-ray counts ($0.5 -
2.0$ keV). Following Szokoly et al. (\cite{Szokoly04}) when $H$ is an upper
limit we set $HR = -1$, while when $S$ is an upper limit we set $HR = 1$. The
14 sources with upper limits in both $H$ and $S$ bands were excluded.  Szokoly
et al. (\cite{Szokoly04}) have shown that absorbed, type 2 AGN are
characterized by $HR \ge -0.2$. We adopt this criterion and identify those
sources which have $HR \ge -0.2$ as absorbed sources. We find 294 (CDF-S: 104,
HDF-N: 190) such absorbed sources which represent $35^{+3}_{-2}\%$ ($1\sigma$
Poisson errors are from Gehrels \cite{Gehrels86}) of the X-ray sources in the
Alexander catalogues. The hardness ratio distribution is shown in Fig.
\ref{HR}. We note that for an $\alpha_{\rm x} = 0.8$ the flux in the $2 - 8$
keV band used by Alexander et al. (\cite{Alexander03}) is only $\sim 15\%$
smaller than that in the commonly used $2 - 10$ keV band. We also point out
that increasing redshift makes the sources softer (e.g., at $z = 3$ the
rest-frame $2 - 8$ keV band shifts to $0.5 - 2$ keV) so our selection
criterion will mistakenly discard some high-z type 2 sources, as pointed out
by Szokoly et al. (\cite{Szokoly04}). The number of type 2 candidates we find
has therefore to be considered a lower limit.

One commonly adopted definition of absorbed source is $N_{\rm H} > 10^{22}$
cm$^{-2}$, where $N_{\rm H}$ is the intrinsic absorption at the redshift of
the source. We discuss below how this compares with our definition. 

\subsection{Finding the optical counterparts}

The optical counterparts to the X-ray sources were selected by
cross-matching the absorbed X-ray sources with the GOODS ACS catalogues
(29\,599 sources in the CDF-S, 32\,048 in the HDF-N). The GOODS catalogues
contain sources that were detected in the $z$-band, with $BVi$ photometry
in matched apertures (Giavalisco et al. \cite{Giava04b}). We checked for
possible offsets between the optical and X-ray astrometry by cross-matching
the full (CDF-S and HDF-N) Alexander catalogues with the GOODS catalogues
using a threshold radius of 1.25\arcsec (see below). In the HDF-N we
detected a systematic shift of $-0.029$\arcsec~in R.A., and $-0.297$\arcsec~in
declination of the Alexander positions with respect to the GOODS positions,
and we applied this correction to the Alexander coordinates before doing
the cross-match.

Note that the ACS image areas for the CDF-S and the HDF-N are both smaller
than (and are completely within) their respective Chandra image fields. 546
out of the 829 sources in the Alexander catalogue (CDF-S: 222/326, HDF-N:
324/503) fall within the ACS image area. Of the 294 absorbed sources, 203
(CDF-S: 77, HDF-N: 126) fall within the ACS image area.

To find the optical counterparts of the absorbed X-ray sources, we initially
search for optical sources that lie within a relatively large threshold radius
of 3.5\arcsec\ (corresponding to the maximal $3\sigma$ positional uncertainty
of the X-ray positions) around each X-ray source. This was done using the
cross match facility in the AVO prototype tool using the ``best match'' mode.

We find 195 ``best'' matches, 76 in the CDF-S and 119 in the HDF-N. So, in the
CDF-S, all but one of the absorbed sources in the ACS image area have optical
counterparts. The remaining source, Alexander ID 213, falls within the
10\arcsec\ disk of a bright galaxy. In the HDF-N there are there are 7
absorbed sources within the image area that do not have an optical counterpart
(these 7 sources do however have optical matches in Barger et al.
(\cite{Barger03}), although we note that most of these are below the 5$\sigma$
limit of their catalogue).

Since the 3.5\arcsec\ radius is large relative to the median positional error,
and given the optical source density the initial cross match inevitably
includes a number of false and multiple matches. To limit our sample to good
matches, we use the criterion that the cross match distance be less than the
combined optical and X-ray $3\sigma$ positional uncertainty for each
individual match. Applying this distance/error $<1$ criterion we limit the
number of matches to 168 (CDF-S: 65, HDF-N: 103). These matches are all within
a much smaller radius than our initial 3.5\arcsec\ threshold, with most of the
distance/error $<1$ matches being within 1.25\arcsec~(and two matches at
1.4 and 1.5\arcsec).

Considering not only the best match but also the multiple matches
within a threshold radius of 1.25\arcsec\, (and distance/error $<1$)
we find 189 matches (CDF-S: 67, HDF-N: 122) to the 168 X-ray
sources. This means that our method of only considering the best
matches does discard 21 possibly valid matches in preference for a
closer match.

The number of false matches we expect to have, as detailed in Appendix
\ref{app1}, is small, between 8 and 15\%.

Our method requires $R$ magnitudes in order to estimate the X-ray luminosity
via the Fiore et al. correlation. The ACS band closest to the $R$ band is the
$i$ band. We then convert the ACS $i$ magnitudes to the R band assuming
$(R-i_{ACS}) = 0.5$, which is the typical value we derive for previously known
type 2 AGN in the GOODS fields (see Sect. \ref{prev}). The $R$ band flux,
$f(R)$, was then computed by converting $R$ magnitudes to specific fluxes and
then multiplying by the width of the $R$ filter (Zombeck 1990), as in Fiore et
al. (\cite{Fiore03}).

\subsection{Estimating the X-ray power}\label{estimate}

As discussed above, previously classified sources and their spectroscopic
redshifts are available from two catalogues: Szokoly et al. (\cite{Szokoly04})
for the CDF-S and Barger et al. (\cite{Barger03}) for the HDF-N. For these
sources we derived the $2 - 8$ keV X-ray power, $L_{2 - 8}$, using our adopted
cosmology and assuming $\alpha_{\rm x} = 0.8$ for the k-correction (Fiore et
al.  \cite{Fiore03}). No absorption correction was applied, which means that
the intrinsic powers of our sources could be larger. As we are dealing with
the hard X-ray band, however, this is probably not going to make much
difference, as pointed out by Fiore et al. (\cite{Fiore03}).

The known sources in the HDF-N were easily identified in our list of
candidates because Barger et al. (\cite{Barger03}) used the same Alexander et
al.  (\cite{Alexander03}) catalogue of X-ray sources. All 103 HDF-N sources
that passed the distance/error $<1$ criterion are listed in Barger et
al. (\cite{Barger03}), 54 of which have spectroscopic redshifts, leaving 47
unclassified HDF-N sources. For the CDF-S Szokoly et al. (\cite{Szokoly04})
utilised the X-ray catalogue of Giacconi et al. (\cite{Giacconi02}) [which is
derived from the same data as Alexander et al. (\cite{Alexander03}) but with
somewhat different source extraction procedures], and list only the positions
of their optical counterparts as determined from VLT FORS1 R-band imaging. To
establish which CDF-S candidates are already listed in the Szokoly et
al. (\cite{Szokoly04}) catalogue we cross matched the GOODS ACS optical
positions of our candidates with the Szokoly optical positions, and inspected
all cases where there were multiple or possibly confused matches. Out of the
65 CDF-S sources which passed the distance/error $ <1$ criterion, 54 are
listed in Szokoly et al. (\cite{Szokoly04}), and 44 of these have
spectroscopic redshifts. This leaves a total of 68 (CDF-S: 21, HDF-N: 47)
unclassified candidates.

For the unclassified sources we estimated the X-ray power as follows: we first
derived the $f(2 - 10 keV)/f(R)$ flux ratio, and then estimated the X-ray
power from the correlation found by Fiore et al. (\cite{Fiore03}), namely
$\log L_{2 - 10} = log f(2 - 10 keV)/f(R) + 43.05$ (Fiore, p.c.; see their
Fig. 5).  Note that this correlation has an r.m.s. of $\sim 0.5$ dex in X-ray
power and that, since the X-ray powers in the Fiore et al. (\cite{Fiore03})
correlation have been corrected for absorption, the estimated powers are
already automatically corrected.

We stress that our estimated X-ray powers reach $\sim 10^{45}$ erg/s and
therefore fall within the range of the Fiore et al. (\cite{Fiore03})
correlation. 


         
\subsection{Finding the type 2 AGN candidates}\label{find} 

The work of Szokoly et al. (\cite{Szokoly04}) has shown that absorbed, type 2
AGN are characterized by $HR \ge -0.2$. It is also well known that normal
galaxies, irrespective of their morphology, have X-ray powers that reach, at
most, $L_{\rm x} \la 10^{42}$ erg/s (e.g., Forman, Jones \& Tucker
\cite{Forman94}, Shapley, Fabbiano \& Eskridge \cite{Shapley01}, Cohen
\cite{Cohen03}). Therefore, any X-ray source with $HR \ge -0.2$ and $L_{\rm x}
> 10^{42}$ erg/s should be an obscured AGN. Furthermore, following Szokoly et
al. (\cite{Szokoly04}), any such source having $L_{\rm x} > 10^{44}$ erg/s
will qualify as a type 2 QSO.

It is a well-established fact that, at a given optical magnitude, galaxies are
also weaker X-ray emitters than AGN, that is they have much smaller
X-ray-to-optical flux ratios, typically $\la 0.1$ (Maccacaro et al.
\cite{Maccacaro88}). Indeed, the Fiore et al. (\cite{Fiore03}) correlation
implies that $L_{\rm x} < 10^{42}$ erg/s corresponds to $f(2 - 10 keV)/f(R) <
0.1$, so a selection based on X-ray-to-optical flux ratio would produce the
same results.


\section{Results}\label{results}

   \begin{table*}
      \caption{Type 2 AGN Candidates, HDF-N.}
   \centering
   \hspace{-7cm}
   \vspace{-3cm}
   \includegraphics[width=\textwidth]{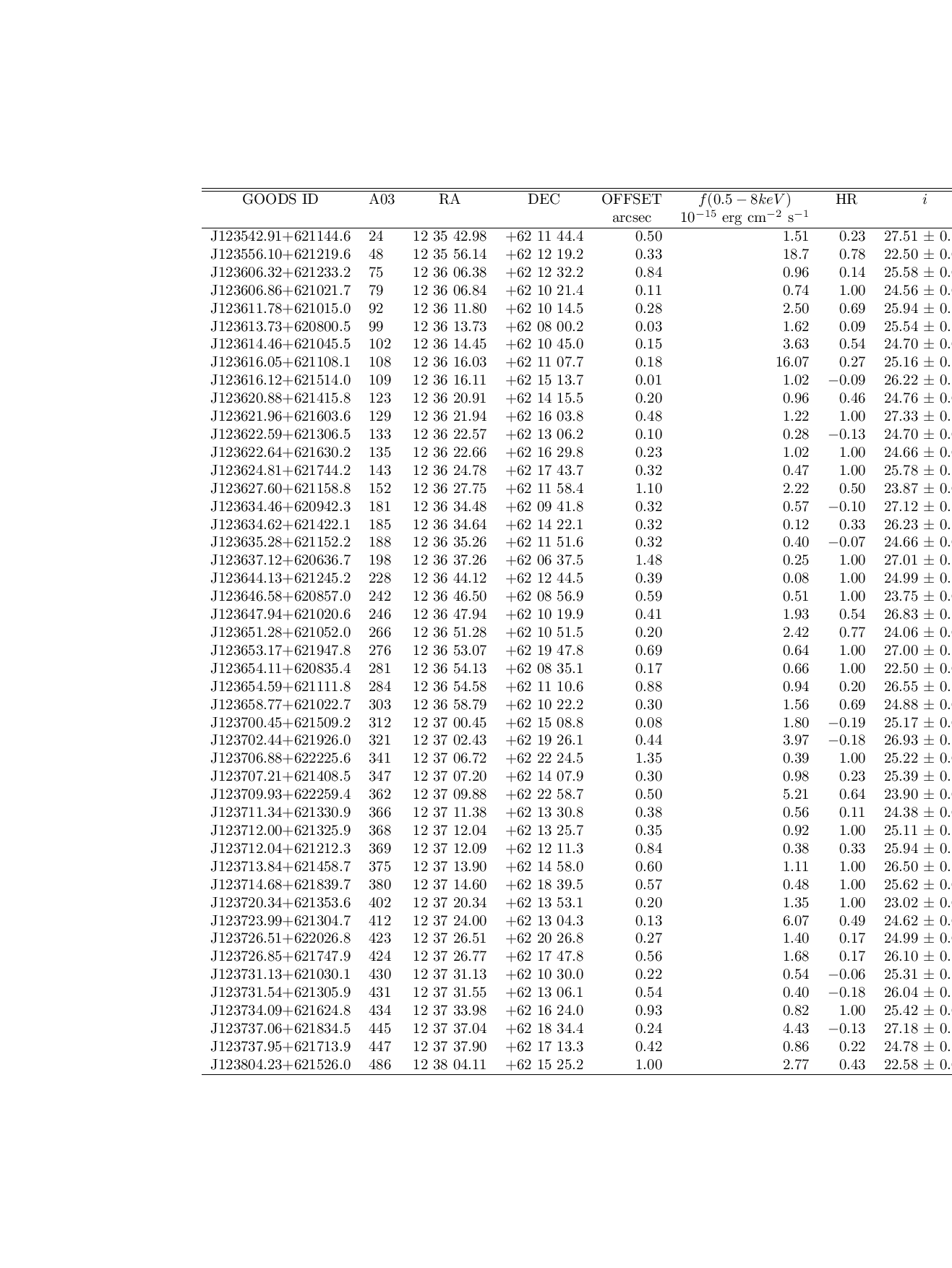}
         \label{tab1}
   \end{table*}

   \begin{table*}
      \caption{Type 2 AGN Candidates, HDF-S.}
   \centering
   \hspace{-8cm}
   \vspace{-10cm}
   \includegraphics[width=\textwidth]{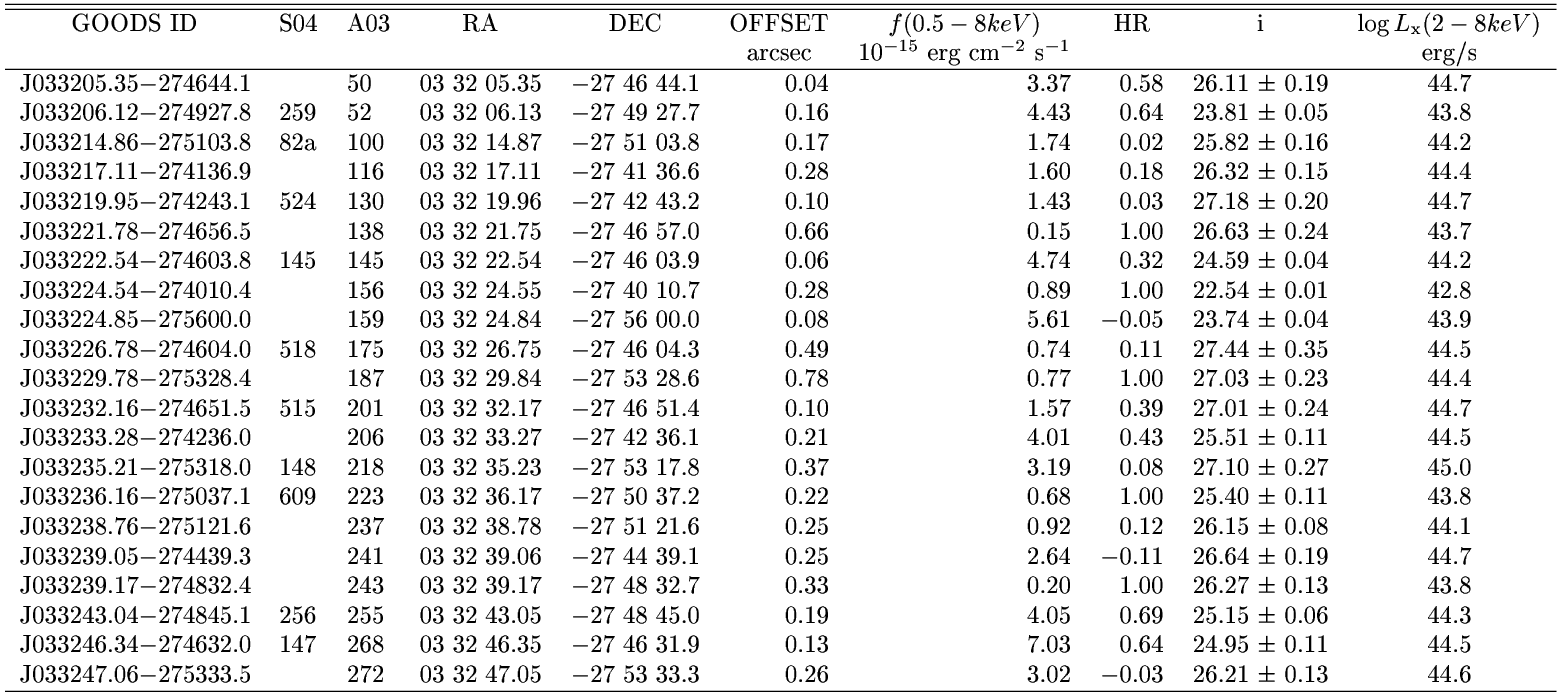}
      \vspace{2cm}
         \label{tab2}
   \end{table*}

Our selection criteria have produced 68 type 2 AGN candidates, based on the
definitions and the method described above. These sources are listed in
Tab. \ref{tab1} (HDF-N) and \ref{tab2} (CDF-S). Table \ref{tab1} gives the
GOODS ID in column (1), the Alexander et al. (\cite{Alexander03}) IDs in
column (2), the X-ray position in columns (3)-(4), the offset between the
X-ray and optical positions in arcseconds in column (5), the full-band ($0.5 -
8$ keV) X-ray flux in column (6), the hardness ratio in column (7), the ACS
$i$ magnitude in column (8), the estimated X-ray power in column (9). Tab.
\ref{tab2} gives, in addition, the Szokoly al. (\cite{Szokoly04}) IDs in
column (2), and the remaining columns are then shifted by one.

Out of our candidates, 31 satisfy the further requirement $L_{2 - 10} >
10^{44}$ erg/s and therefore qualify as QSO 2 candidates. We note that the
distribution of estimated X-ray power covers the range $5 \times 10^{42} - 2
\times 10^{45}$ erg/s and peaks around $10^{44}$ erg/s (see Fig. \ref{LX},
dashed line). The number of QSO 2 candidates, therefore, is very sensitive to
the dividing line between low- and high-luminosity AGN, which is clearly
arbitrary and cosmology dependent. For example, if one defines as QSO 2 all
sources with $L_{2 - 10} > 5 \times 10^{43}$ erg/s, a value only a factor of 2
below the commonly used one and corresponding to the break in the AGN X-ray
luminosity function (Norman et al. \cite{Norman02}), the number of such
sources increases by $\sim 50\%$. We also note that, based on the r.m.s.
around the Fiore et al. (\cite{Fiore03}) correlation, the number of QSO 2
candidates fluctuates in the $13 - 54$ region. The number of type 2 AGN, on
the other hand, can only increase, as all our candidates have estimated $\log
L_{2 - 10} > 42.5$.

It is interesting that all candidates with $HR \ge -0.2$ have estimated X-ray
power $L_{\rm x,est} > 10^{42}$ erg/s, and therefore all qualify as AGN. Some
previously known sources, however, do have X-ray powers below this value (see
Sect. \ref{prev}).

As expected, being still unidentified, our sources are very faint: their
median ACS $i$ magnitude is $\sim 25.5$, which corresponds to $R \sim 26$
(compare this to the $R \sim 22$ typical of the CDF-S sources with redshift
determination). The QSO 2 candidates are even fainter, with median $i$
magnitude $\sim 26.3$ ($R \sim 26.8$).

It is important to notice that 6 of the 21 sources in Tab. \ref{tab2} have
actually been observed by Szokoly et al. (\cite{Szokoly04}; see their Fig. 5)
but either their spectrum is featureless (4 sources) or the continuum is so
weak that no extraction could be made (2 sources).
 
\subsection{Testing our estimated X-ray powers}\label{testlx}

To check the reliability of our estimated X-ray powers we compared the
predicted and observed luminosities for the already known sources. The Fiore
et al. (\cite{Fiore03}) correlation is valid for non-type 1 AGN and galaxies
so we did the comparison for these sources. Namely, we excluded the type 1
sources from the Szokoly et al. (\cite{Szokoly04}) catalogue and the sources
with broad lines (type ``B'') in Barger et al. (\cite{Barger03}). Moreover, we
only considered sources with quality flag $\ge 2.0$ in the former catalogue
and excluded objects with less secure redshifts (type ``s'') in the latter, to
have reliable redshifts and therefore powers. This leaves us with 122
objects. For these, we find that $L_{\rm x,est} \propto L_{\rm
x}^{0.93\pm0.06}$, where the slope is the mean ordinary least-squares slope
(Isobe et al. \cite{Isobe90}). The mean values of the estimated ($\langle \log
L_{\rm x,est}\rangle = 42.57\pm0.08$) and observed ($\langle \log L_{\rm x}
\rangle = 42.49\pm0.09$) luminosities, are consistent. This shows that our
method works for non-type 1 AGN and can then be safely applied to optically
unidentified but absorbed sources.

It is tempting to use our estimated X-ray powers together with the observed
fluxes to derive redshifts for our type 2 candidates. We can first check how
reliable these are by comparing them with the spectroscopic redshifts
available for the known non-type 1 AGN described above. We follow standard
procedures (see, e.g., Mobasher et al. \cite{Mobasher04}) and quantify the
reliability of the estimated redshifts, $z_{\rm est}$, by measuring the
fractional error for each galaxy, $\Delta \equiv (z_{\rm est} - z_{\rm
spec})(1+z_{\rm spec})$. We find a median error, $\langle \Delta \rangle =
0.005$, an r.m.s. scatter, $\sigma(\Delta) \sim 0.29$, and a rate of
``catastrophic'' outliers, $\eta$, defined as the fraction of the full sample
that has $|\Delta| > 0.2$, $\sim 19\%$. Clipping these outliers gives $\langle
\Delta \rangle = -0.03$ and $\sigma(\Delta) \sim 0.1$. For comparison,
Mobasher et al. (\cite{Mobasher04}) have used extensive multiwavelength
photometric data to estimate photometric redshifts for a sample of 434
galaxies with spectroscopic redshifts in the CDF-S. They find $\langle \Delta
\rangle = -0.01$, $\sigma(\Delta) \sim 0.1$, and $\eta \sim 2.4 - 10\%$,
depending on the subsample considered. Excluding the outliers they find values
of $\sigma(\Delta) \sim 0.05$. Given the simplicity of our method, these
results are certainly encouraging and show that our estimates should at least
provide a rough idea of the redshifts at which we might expect our sources to
be.

Applying this method to our type 2 AGN candidate list we find a mean of
$\langle z_{\rm est} \rangle \sim 2.9$ (median 2.6). The QSO~2 have,
as expected, significantly higher estimated redshifts $\langle z_{\rm est}
\rangle \sim 3.7$ compared to $\langle z_{\rm est} \rangle \sim 2.2$ for the
non-QSO candidates.

We can also use the ACS four band photometry to constrain the redshifts of our
sources. Cristiani et al. (\cite{Cristiani04}) have discussed a set of four
optical criteria using the $i-z$, $B-V$, and $V-i$ colours (a variation of the
``B-dropout'' technique) to select AGN in the redshift range $3.5 \la z \la
5.2$. By applying the same criteria to our candidates we find 9 such
sources. Their average estimated redshift is $\langle z_{\rm est} \rangle \sim
3.6$, and all but one sources have $2.8 < z_{\rm est} < 5.4$. Again, this 
shows that our estimated redshifts are, overall, relatively robust.

\begin{table}
\begin{center}
\caption{B dropouts.\label{tab_Bdrop}}
\begin{tabular}{lcrrr}
\hline\hline
GOODS ID & $i$ & $i-z$ & $V-i$ & $B-V$\\
\hline
J033226.78$-$274604.0 &  27.44 &    0.37    & 3.21 &    $-1.36$ \\
J033232.16$-$274651.5 &  27.01 &    0.27    & 0.43 &     1.25\\
J033238.76$-$275121.6 &  26.15 &    0.04    & 0.23 &     1.41\\
J033239.05$-$274439.3 &  26.64 &    0.26    & 0.23 &     4.41\\
J123611.78$+$621015.0 &  25.94 &    0.71    & 2.46 &     $>3.0$\\
J123627.59$+$621158.8 &  23.86 &    0.24    & 0.69 &     1.25\\
J123634.46$+$620942.3 &  27.12 &    0.28    & 1.53 &     $>3.0$\\
J123714.68$+$621839.7 &  25.62 &    0.13    & 0.44 &     1.87\\
J123731.54$+$621305.9 &  26.04 &   $-0.13$  & 0.46 &     1.39\\
\hline
\end{tabular}
\end{center}
\end{table}

These sources are listed in Table \ref{tab_Bdrop}, which gives the GOODS ID in
column (1), the ACS $i$ magnitude in column (2), followed by the $i-z$, $V-i$
and $B-V$ colours (all in the AB system) in columns (3)-(5).

\begin{figure*}
   \centering
HERE FIG. 2
      \caption{CDF-S Cutouts. $B, V, i$, and $z$ ACS image cutouts are
displayed for all the type 2 AGN candidate in the CDF-S. Each image is
3\arcsec $\times$ 3\arcsec in size, with north up and east to the left. The
Alexander identification number is shown in the top left of each panel, and
the circles drawn on the $i$-band images indicate the Alexander X-ray
positions, with the radius indicating the 90$\%$ positional error. The (black
and white) square symbols indicate the positions of the GOODS source which
have been matched to the X-ray positions. The X symbols indicate the positions
of any corresponding sources in the Szokoly catalogue.}
         \label{CDFS_cutouts}
   \end{figure*}

   \begin{figure*}
   \centering
HERE FIG. 3
      \caption{CDFN Cutouts. $B, V, i$, and $z$ ACS 
image cutouts are displayed for all the type 2 AGN candidate in the CDFN.
Each image is 3\arcsec $\times$ 3\arcsec in size, with north up 
and east to the left. The Alexander identification number is
shown in the top left of each panel, and the circles drawn on 
the $i$-band images indicates the Alexander X-ray positions, with 
the radius indicating the $90\%$ positional error. 
The (black and white) square symbols indicate the positions of the 
GOODS sources which have been 
matched to the X-ray positions. 
}
         \label{CDFN_cutouts}
   \end{figure*}

\subsection{Testing our method}

Out of the 9 QSO 2 previously known (see Sect. \ref{disc}), our method has
rediscovered them all as type 2 AGN, 8/9 with $L_{\rm x,est} > 8 \times
10^{43}$ erg/s (the ninth one having $L_{\rm x,est} \sim 4 \times 10^{42}$
erg/s). Out of the 29 type 2 AGN with $L_{2 - 8} > 10^{42}$ erg/s in the
Szokoly et al. (\cite{Szokoly04}) catalogue, 28 have $L_{\rm x,est} > 10^{42}$
erg/s and therefore fulfil our selection criteria, while the last one is
barely below with $L_{\rm x,est} \sim 7 \times 10^{41}$ erg/s.

During the completion of this work three redshift databases covering the CDF-S
and HDF-N became public. Namely: the ESO/GOODS FORS2 spectroscopy data
(Vanzella et al. \cite{Vanzella04}\footnote{\tt
http://archive.eso.org/wdb/wdb/vo/goods/form}), the VIMOS VLT Deep Survey
(VVDS; Le F\`evre et al. \cite{Lefevre04}\footnote{\tt
http://cencosw.oamp.fr/}) and the Team Keck Treasury Redshift Survey (TKRS;
Wirth et al. \cite{Wirth04}\footnote{\tt
http://www2.keck.hawaii.edu/science/tksurvey/}). We cross-matched our
candidate list with these surveys using the AVO prototype, finding only one
match, namely \object{GOODSJ123556.10+621219.6}, with a quite featureless
spectrum and redshift $z = 0.9585$ (our estimated redshift based on its X-ray
power and flux is $z = 0.93$). Given the limits of the three surveys, that is
$z_{\rm AB} <24.5$, $I_{\rm AB} \le 24$, and $R \le 24.4$ respectively, this
is not surprising. The only match we found, in fact, is the brightest of our
northern candidates.

\subsection{Image cutouts} 

Image cutouts of the ACS $BViz$ data centred on each of the new AGN type 2
candidates were generated using the AVO prototype. This was done by sending
requests to the Aladin image server which remotely extracts the required image
sections in the four bands from a database containing the GOODS data. The
hierarchical description of the ACS images in terms of the IDHA data model,
and advanced protocols for sending detailed requests to the server makes this
process more efficient and flexible than traditional image servers.
Figs.~\ref{CDFS_cutouts} and \ref{CDFN_cutouts} display
3\arcsec$\times$3\arcsec\ sections of the cutouts in all four bands for the
CDF-S and HDF-N respectively. Each image is shown with a linear grey-scale
over the range of minimum to maximum image values of the individual
3\arcsec$\times$3\arcsec\ image sections. The 90\% error circle of the
Alexander et al. (\cite{Alexander03}) catalogue X-ray positions is overlaid on
the $i$ band images (since our method used the $i$ band), along with the GOODS
catalogue position (black/white squares). The Szokoly et al.
(\cite{Szokoly04}) optical positions (X symbols) are also shown for the CDF-S
sources (Fig.~\ref{CDFS_cutouts}). Note that in a few cases the position of
the GOODS optical counterpart is outside the X-ray 90\% error circle. These
are still valid matches in terms of the distance/error parameter (see
Sect. \ref{find}), which limits matches to within the $3\sigma$ total
positional uncertainty which is $\sim1.8$ times larger than the 90\% radius.

Our sources cover a large spectrum of morphologies, ranging from extended,
low-surface brightness to point-like. It is interesting to note, however, that
all QSO 2 candidates are point-like. 


\subsection{Hubble Ultra Deep Field data}

\begin{table*}
\caption{Type 2 AGN candidates, UDF}\label{udf_tab}
\begin{tabular}{lllllllc}
\hline\hline
  \multicolumn{1}{c}{S04} &
  \multicolumn{1}{c}{A03} &
  \multicolumn{1}{c}{UDF ID} &
  \multicolumn{1}{c}{RA} &
  \multicolumn{1}{c}{DEC} &
  \multicolumn{1}{c}{OFFSET} &
  \multicolumn{1}{c}{i} &
  \multicolumn{1}{c}{$\log L_{\rm x}(2 - 8 keV)$} \\
  \multicolumn{1}{c}{} &
  \multicolumn{1}{c}{} &
  \multicolumn{1}{c}{} &
  \multicolumn{1}{c}{} &
  \multicolumn{1}{c}{} &
  \multicolumn{1}{c}{arcsec} &
  \multicolumn{1}{c}{} &
  \multicolumn{1}{c}{erg/s} \\
\hline
  515 & 201 & 7326 & 03 32 32.17 & $-$27 46 51.49 & 0.10 & 26.869 $\pm$ 0.021 & 44.6\\
      & 243 & 1441 & 03 32 39.19 & $-$27 48 32.87 & 0.31 & 28.328 $\pm$ 0.046 & 44.7\\
  256 & 255 & 1025 & 03 32 43.04 & $-$27 48 45.08 & 0.18 & 25.153 $\pm$ 0.006 & 44.3\\
\hline\end{tabular}
\end{table*}

   \begin{figure}
   \centering
\includegraphics[width=5.8cm]{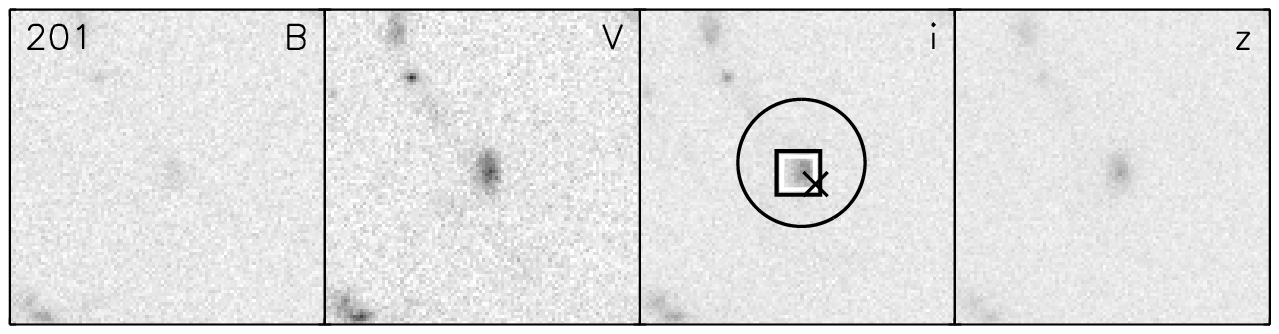}
\includegraphics[width=5.8cm]{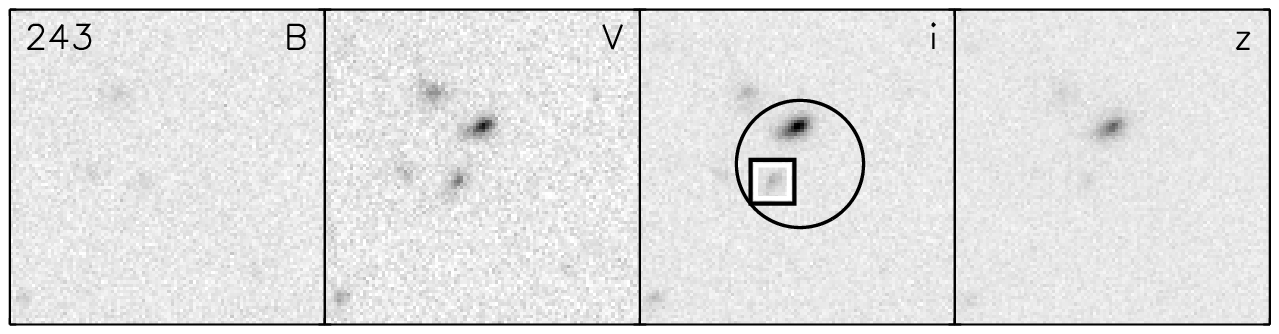}
\includegraphics[width=5.8cm]{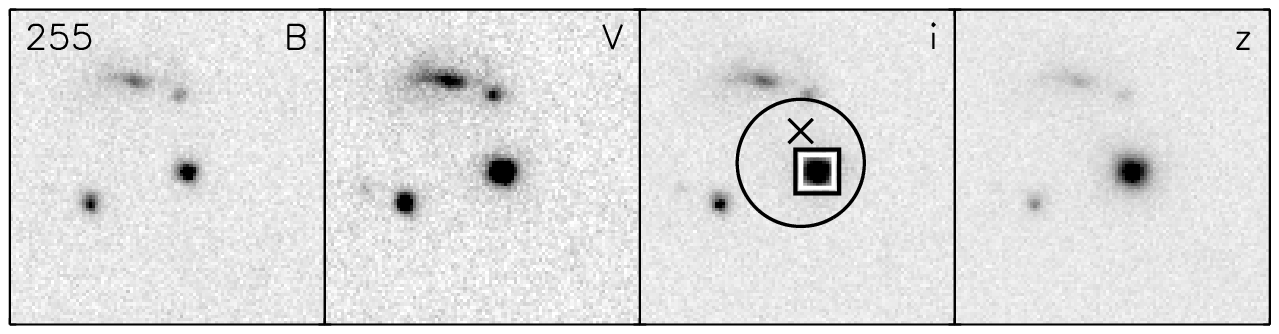}
      \caption{CDF-S UDF Cutouts. $B, V, i$, and $z$ ACS image
cutouts are displayed for the type 2 AGN candidates that
fall within the area of the UDF. The (black and white) square 
symbols indicate the positions of the UDF sources in the 
UDF catalogue.The X symbols indicate the positions
of any corresponding sources in the Szokoly catalogue.}
         \label{CDFS_udf_cutouts}
   \end{figure}

A number of sources also fall within the Hubble Ultra Deep Field (UDF;
Beckwith et al. \cite{Beckwith04}). The UDF provides extremely deep ACS Wide
Field Channel $BViz$ imaging (proposal IDs 9\,978 and 10\,086) of a single
5.25\arcmin $\times$ 5.25\arcmin\ field centred on 03$^h$32$^m$39.0$^s$,
-27\degr47\arcmin29.1\arcsec\ (J2000) which lies within the GOODS CDF-S
field. Five of the absorbed sources fall within this field, and 3 of these
(A03 201, 243 and 255) are in our list of unclassified type 2 AGN candidates
(Tab. \ref{tab2}). For these sources we cross matched the X-ray positions with
the UDF Version 1.0 $BViz$ $i$-band detected catalogues\footnote{\tt
http://www.stsci.edu/hst/udf} to find the UDF optical counterparts. We then
used the UDF $i$ band magnitudes to estimate the X-ray power using the Fiore
et al. (\cite{Fiore03}) correlation as described above. The estimated X-ray
powers, and UDF $i$ band magnitudes for these 3 sources are listed in Table
\ref{udf_tab}\, and UDF image cutouts overlaid with the UDF, Alexander et
al. (\cite{Alexander03}), and Szokoly et al. (\cite{Szokoly04}) positions are
shown in Fig.~\ref{CDFS_udf_cutouts}. Note that for one source, A03 243, we
find a different, fainter, optical counterpart than was found using the GOODS
catalogue, resulting in a larger estimated X-ray power. Classically, one would
have solved this dilemma by obtaining a spectrum of both sources. However,
given the faintness of the two candidates ($i \sim 26.3$ and 28.3), this is
unlikely to happen in the near future. [The two previously classified sources
that fall within the UDF are A03 245 (QSO-2) and 242 (AGN-1)].

Note that none of our candidates have spectra obtained with
the ACS High Resolution Channel grism\footnote{\tt
http://www.stecf.org/UDF/goodsdata.html} as part of the UDF observing
campaign.

\section{Discussion}\label{disc}

\subsection{Previously known type 2 AGN}\label{prev}

We defined a sample of previously known type 2 AGN in the GOODS ACS fields as
follows. In the HDF-N, we selected absorbed ($HR \ge -0.2$) sources without
broad lines (type ``B'' in Barger et al. \cite{Barger03}) having X-ray power
$L_{2 - 8} > 10^{42}$ erg/s (see Sect. \ref{find}). In the CDF-S we initially
selected sources with ``AGN-2'' or ``QSO-2'' classifications in Szokoly et
al. (\cite{Szokoly04}). We then applied the further criteria that $L_{2 - 8} >
10^{42}$ erg/s and $HR \ge -0.2$ in order to be consistent with the HDF-N
sample. Successive power and $HR$ cuts exclude 2 and 3 sources respectively,
with a further 3 sources excluded because they have no counterpart in the
Alexander et al. (\cite{Alexander03}) catalogue (which is necessary for
consistent definitions of $HR$ and $L_{2 - 8}$). [Note that differences
between our $HR$ values and those listed by Szokoly et al. (\cite{Szokoly04})
can be explained by the different input catalogues produced by Giacconi et al.
(\cite{Giacconi02}) and Alexander et al. (\cite{Alexander03})]. These
definitions lead to a total of 79 known type 2 AGN (CDF-S: 35, HDF-N:
44). These are listed in Tables \ref{tab_known-N} and \ref{tab_known-S}, which
give the Alexander et al. (\cite{Alexander03}) and Szokoly et al.
(\cite{Szokoly04}) (for the CDF-S) IDs, redshift, hard X-ray power, and
notes. Only 9 of these sources can be classified as QSO 2 ($L_{2 -
8} > 10^{44}$ erg/s), 3 of which have poor redshift determinations (two
sources with type ``s'' flags in Barger et al. \cite{Barger03}, and one source
with quality flag $= 1$ in Szokoly et al. \cite{Szokoly04}).  We also note
that with a lower X-ray power cut of $L_{2 - 8} > 5 \times 10^{43}$ erg/s the
number of QSO 2 increases to 13.

\begin{table}
\begin{center}
\caption{Previously Known Type 2 AGN, HDF-N.\label{tab_known-N}}
\begin{tabular}{rrccl}
\hline\hline
A03 & redshift & $\log L_{\rm x}(2 - 8 keV)$ &
Notes$^a$\\
  &   & erg/s &   \\
\hline
 72 & 0.936 &  42.73 &    m \\
 76 & 0.637 &  42.71 &    m\\
 82 & 0.681 &  42.52 & \\
 83 & 0.459 &  42.41 & \\
 90 & 1.140 &  43.33 &\\
106 & 1.017 &  42.31 &\\
121 & 0.520 &  42.27 &\\
122 & 1.338 &  42.90 &    s\\
127 & 1.014 &  43.27 &    m\\
142 & 0.747 &  42.47 &\\
150 & 0.762 &  42.82 &\\
157 & 1.264 &  43.89 &\\
158 & 1.013 &  43.02 &\\
160 & 0.847 &  42.90 &    m\\
163 & 0.485 &  42.72 &\\
164 & 0.953 &  42.92 &\\
170 & 0.680 &  42.44 &\\
171 & 1.995 &  43.63 &    m\\
187 & 0.847 &  42.27 &\\
190 & 2.005 &  43.76 &\\
191 & 0.560 &  42.79 &    m\\
201 & 1.020 &  42.74 &    m\\
217 & 0.518 &  42.15 &\\
229 & 1.487 &  42.23 &    s\\
240 & 0.961 &  43.95 &\\
259 & 1.609 &  44.48 &    m\\
261 & 0.902 &  42.15 &\\
267 & 0.401 &  42.15 &\\
278 & 1.023 &  42.92 &\\
330 & 3.406 &  43.43 &\\
352 & 0.936 &  42.47 &\\
373 & 0.475 &  42.25 &\\
384 & 1.019 &  43.45 &\\
390 & 1.146 &  44.04 &    s\\
398 & 2.638 &  44.05 &    s\\
405 & 0.978 &  42.96 &\\
409 & 2.240 &  43.61 &\\
413 & 0.474 &  42.60 &\\
439 & 0.636 &  42.36 &\\
441 & 0.634 &  42.24 &\\
442 & 0.852 &  42.33 &\\
448 & 1.238 &  43.32 &\\
462 & 0.511 &  42.03 &\\
468 & 0.911 &  42.39 &\\
\hline
\end{tabular}
\end{center}
{\footnotesize{$^a$From Barger et al. \cite{Barger03}: s,
less secure redshift identification based primarily on a single emission line
and the continuum shape; m, complex or multiple structure or possible
contamination of the photometry by a nearby brighter object.}}
\end{table}

\begin{table}
\begin{center}
\caption{Previously Known Type 2 AGN, CDF-S.\label{tab_known-S}}
\begin{tabular}{rrccl}
\hline\hline
A03 & S04 & redshift & $\log L_{\rm x}(2 - 8 keV)$ &
Notes$^a$\\
  &   &   & erg/s &   \\
\hline
 44 &      66 &       0.574 &   43.31 &\\
 48 &     267 &       0.720 &   43.07 & QF=1\\
 60 &     155 &       0.545 &   42.11 &\\
 68 &      62 &       2.810 &   44.48 &\\
 80 &     535 &       0.575 &   42.20 &\\
 84 &     149 &       1.033 &   42.81 & QF=1\\
 86 &      57 &       2.562 &   44.28 &\\
 88 &     56a &       0.605 &   43.41 &\\
 90 &     600 &       1.327 &   42.86 &\\
 91 &     266 &       0.735 &   42.43 &\\
 96 &     531 &       1.544 &   43.19 &\\
118 &     51a &       1.097 &   44.06 &\\
123 &     153 &       1.536 &   43.97 &\\
126 &      50 &       0.670 &   42.61 & QF=1\\
131 &     253 &       0.481 &   42.53 & QF=1\\
134 &     151 &       0.604 &   43.02 &\\
137 &    612b &       0.736 &   42.55 &\\
146 &     188 &       0.734 &   42.14 &\\
160 &    606a &       1.037 &   42.89 & QF=1\\
161 &      47 &       0.733 &   43.08 &\\
162 &     260 &       1.043 &   42.83 &\\
164 &     150 &       1.090 &   43.34 &\\
166 &      45 &       2.291 &   44.20 & QF=1\\
176 &      43 &       0.734 &   42.90 &\\
179 &      41 &       0.668 &   43.31 &\\
188 &     202 &       3.700 &   44.47 &\\
212 &     512 &       0.668 &   42.08 &\\
220 &     190 &       0.735 &   43.07 &\\
245 &      27 &       3.064 &   44.64 &\\
247 &      25 &       0.625 &   43.15 & QF=0.5\\
249 &    611a &       0.979 &   42.74 & QF=1\\
263 &     20a &       1.016 &   43.40 &\\
269 &     170 &       0.664 &   42.38 &\\
271 &     252 &       1.178 &   43.29 &\\
276 &     184 &       0.667 &   42.54 &\\
\hline
\end{tabular}
\end{center}
{\footnotesize{$^a$From Szokoly et al. (\cite{Szokoly04}): QF = 1, spectrum
cannot be identified securely, typically only a single narrow line present; QF = 0.5,
only a hint of some spectral feature.}}
\end{table}

\begin{figure}
  \centering
  \resizebox{\hsize}{!}{\includegraphics{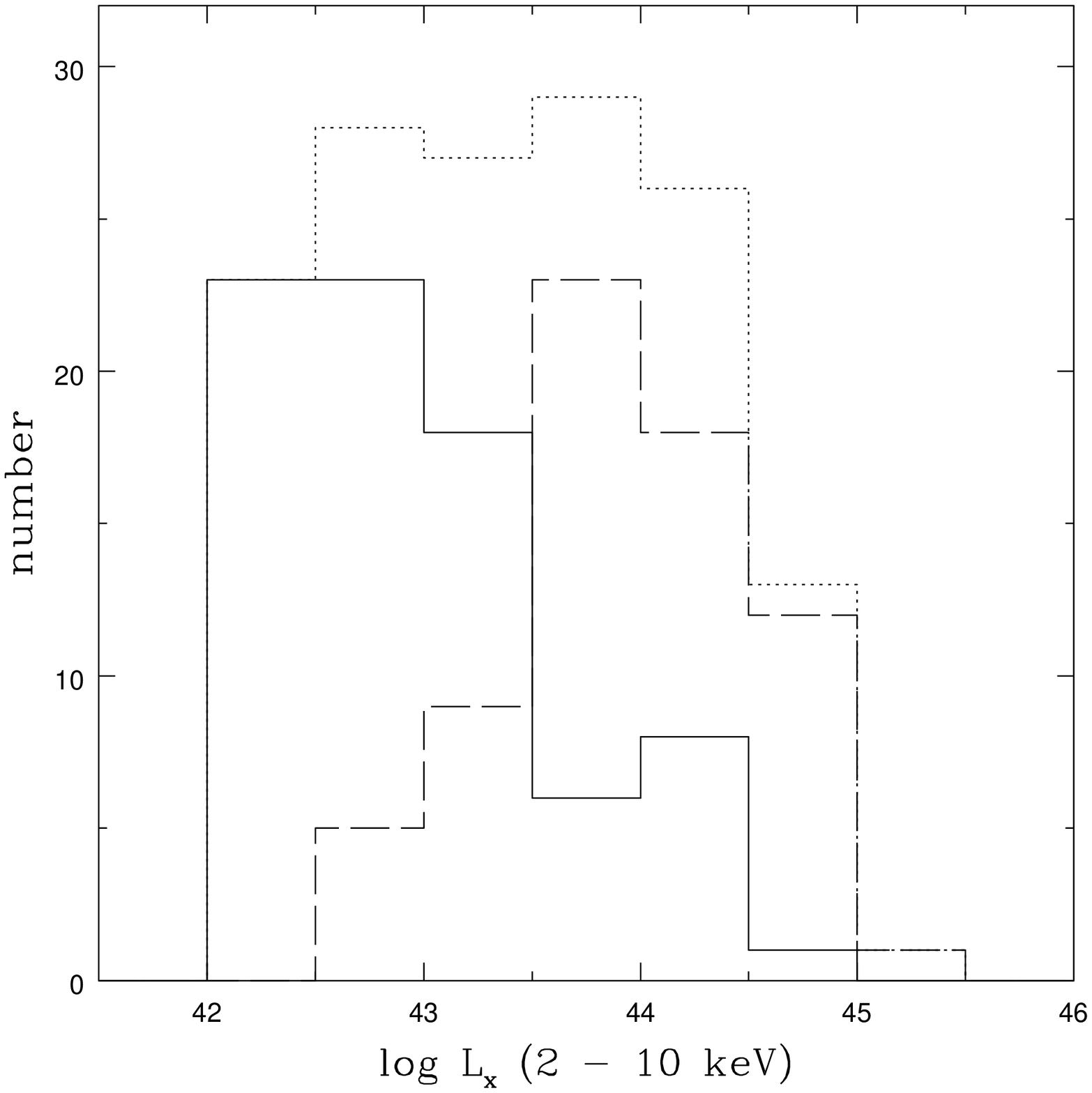}}
     \caption{The X-ray power distribution for our new type 2 AGN candidates
             (dashed line), previously known type 2 AGN (solid line), and
             the sum of the two populations (dotted line). QSO 2 are defined,
             somewhat arbitrarily, as having $L_{\rm 2- 10 keV} > 10^{44}$ erg/s.}
        \label{LX}
  \end{figure}

Fig. \ref{LX} shows the X-ray power distribution for our new type 2 AGN
candidates (dashed line), previously known type 2 AGN (solid line), and the
combined sample (dotted line). It is interesting to note how the distributions
are very different, with the already known type 2 AGN peaking around $L_{\rm
x} \sim 10^{43}$ erg/s and declining for luminosities above $\sim 3 \times
10^{43}$ erg/s, while our new candidates are rising in this range and peak
around $L_{\rm x} \sim 10^{44}$ erg/s. To be more quantitative, while only
$\sim 1/5$ of already known type 2 AGN have $\log L_{\rm x} > 43.5$, $\sim
3/4$ of our candidates are above this value. This difference is easily
explained by our use of the X-ray-to-optical flux ratios to estimate X-ray
powers and by the fact that our candidates are on average $\sim 3$ magnitudes
fainter than previously known sources. Our method is then filling a gap in the
luminosity distribution, which becomes almost constant in the range $10^{42}
\la L_{\rm x} \la 3 \times 10^{44}$ erg/s. This also explains the fact that
the number of QSO 2 candidates we find is $\ga 3$ times larger than the
previously known ones.

Conversely the fluxes of the already known QSO 2 reach $f(2 - 8 keV) \sim 3
\times 10^{-15}$ erg cm$^{-2}$ s$^{-1}$, while our new QSO 2 candidates go
down to the fainter limit of $f(2 - 8 keV) \sim 4 \times 10^{-16}$ erg
cm$^{-2}$ s$^{-1}$. This is explained by the much fainter optical magnitudes
we are probing with our method which, for a given X-ray-to-optical flux ratio,
translate into fainter X-ray fluxes as well.

By going fainter we are also probing the type 2 AGN population at higher
redshifts. While our estimated redshift is $\langle z_{\rm est} \rangle \sim
2.9$ (median 2.6), previously known type 2 AGN have $\langle z \rangle \sim
1.1$ (median 0.9). For QSO 2 we find $\langle z_{\rm est} \rangle \sim 3.7$
(median 3.5), as compared to $\langle z \rangle \sim 2.3$ (median 2.6).

\subsection{Total number of type 2 AGN}\label{type2}

We find a total of 147 type 2 AGN in the GOODS ACS area, 79 of which were
already known. This corresponds to $27^{+3}_{-2}\%$ of the 546 sources in the
Alexander et al. (\cite{Alexander03}) catalogue ($14^{+2}_{-2}\%$ including
the previously known sources only). As regards QSO 2, the total number is 40,
only 9 of which were previously known. Our method has therefore more than
quadrupled the number of such sources in the GOODS ACS fields. These represent
$7^{+1}_{-1}\%$ of all (546) X-ray sources (the previously known QSO 2 make up
$2^{+1}_{-1}\%$).

Note that all of these sources satisfy the commonly adopted definition of
absorbed source discussed above. In fact, Fig. 12 of Szokoly et al.
(\cite{Szokoly04}) shows that $HR \ge -0.2$ corresponds to $N_{\rm H} >
10^{22}$ cm$^{-2}$, for an intrinsic $\alpha_{\rm x} = 1$, for $z \ga
0.4$. The lowest redshift for previously known type 2 AGN is 0.4, while the
lowest estimated redshift for our type 2 candidates is 0.9. The typical $HR$
and redshifts for our sources imply that we are dealing with $N_{\rm H}
\approx 10^{23}$ and $\approx 3 \times 10^{23}$ cm$^{-2}$ for the
known and candidate sources respectively.  

X-ray obscuration can be also present in a small fraction of broad-lined
AGN. Perola et al. (\cite{Perola04}; and references therein) find this to
be the case in $\sim 10\%$ of their sources, with estimated $N_{\rm H} \ga
10^{22}$ cm$^{-2}$. As discussed in the previous paragraph, these
sources are then likely to be at relatively low redshift. Therefore,
contamination by type 1 AGN in our sample is expected to be negligible.

\subsection{The surface density of QSO 2}\label{dens}

\begin{table*}
\begin{center}
\caption{QSO 2 surface density.\label{counts}}
\begin{tabular}{lrrrrrr}
\hline\hline
$f_{\rm} (0.5 - 8 keV)$ & \multicolumn{3}{c}{N} & \multicolumn{3}{c}{$N(>f_{\rm x})$} \\
  & known & candidate & total & known & candidate & total \\ 
erg cm$^{-2}$ s$^{-1}$ & & & & \multicolumn{3}{c}{deg$^{-2}$} \\
\hline
$1 \times 10^{-15}$ & 9 & 25 & 34 & $89^{+40}_{-29}$ & $247^{+60}_{-49}$ & $335^{+68}_{-57}$ \\
$2 \times 10^{-15}$ & 9 & 14 & 23 & $89^{+40}_{-29}$ & $138^{+48}_{-36}$ & $227^{+58}_{-47}$ \\
$5 \times 10^{-15}$ & 7 & 3 & 10 & $69^{+37}_{-25}$ & $30^{+29}_{-16}$ & $99^{+42}_{-31}$ \\
$1 \times 10^{-14}$ & 3 & 1 & 4 & $30^{+29}_{-16}$ & $10^{+23}_{-8}$ & $39^{+31}_{-19}$ \\
\hline
\end{tabular}
\end{center}
\end{table*}

The detection of faint type 2 AGN candidates and the careful assessment of the
previously known such sources in the two GOODS fields allow us to put strong
constraints on the surface density of type 2 sources, and in particular on
that of QSO 2. We use a value for the $Viz$-band coverage of 365 arcmin$^2
\simeq 0.1014$ deg$^2$ (Giavalisco et al. \cite{Giava04}).

Due to the variable sensitivity of the Chandra ACIS-I detector across the field of
view, the area in which fainter X-ray sources could be detected is smaller
than that for brighter sources. In other words, the sky coverage is not
uniform at all fluxes. Fig. 5 of Giacconi et al. (\cite{Giacconi02}) and 
Fig. 19 of Alexander et al. (\cite{Alexander03}), however, show that the effect
is strong only at relatively small fluxes. For example, for both CDF-S and
HDF-N the effective area decreases in the hard band, for which this effect is
strongest, by $\ga 25\%$ only for $f_{\rm} (2 - 8/10 keV) \la 10^{-15}$ erg
cm$^{-2}$ s$^{-1}$.

To better quantify the magnitude of this correction we evaluated the sky
coverage for the combined GOODS fields by simply scaling the two sky coverages
so that the maximum area for each field was equal to half the total GOODS
area, summing then up the two areas. This works only as a first approximation
and slightly overestimates the correction since the GOODS ACS fields are at
the centre of the Chandra fields and therefore deeper than average in the
X-ray band. In other words, the sky coverage should be re-computed as its
shape would change, with a larger area accessible at fainter fluxes and
therefore a smaller correction than we estimate. In any case, under our
assumptions we find that the correction is within the $1\sigma$ Poisson range,
and therefore within the statistical uncertainties, for $f_{\rm} (0.5 - 8 keV)
\ga 10^{-15}$ erg cm$^{-2}$ s$^{-1}$. Our conservative approach is then to not
to take the sky coverage into account and limit ourselves to X-ray fluxes $\ge
10^{-15}$ erg cm$^{-2}$ s$^{-1}$. This means that our values are actually {\it
lower limits}, although the real surface densities should not be more than
$25\%$ larger.

Table \ref{counts} gives the QSO 2 surface density for four flux limits,
splitting the contribution into known and candidate sources for the full $0.5
- 8$ keV band. (The counts in the hard band are not very different, 
since we are dealing with absorbed sources, with the total numbers
changing to 33 [from 34] and 8 [from 10] for the first and third flux limit
respectively.)

These numbers can be compared with recent estimates and predictions. Perola et
al. (\cite{Perola04}) find a surface density of highly obscured QSO, which
they define as having $L_{2 - 10} > 10^{44}$ erg/s and $N_{\rm H} > 10^{22}$
cm$^{-2}$, $\sim 48$ deg$^{-2}$ for $f_{\rm} (2 - 10 keV) > 10^{-14}$ erg
cm$^{-2}$ s$^{-1}$, consistent with our value. At lower fluxes the situation
is different. For example, for $f_{\rm} (0.5 - 7 keV) > 5 \times 10^{-15}$ erg
cm$^{-2}$ s$^{-1}$ Gandhi et al. (\cite{Gandhi04}) quote an estimated value
$>3$ deg$^{-2}$ and possibly higher than 10 deg$^{-2}$ and a predicted value
from their model of 19 deg$^{-2}$. We find a density $\sim 100$ deg$^{-2}$,
that is a factor $\ga 10$ larger than their estimate and $\sim 5$ times larger
than their prediction. Already known sources make up $\sim 70\%$ of our total
value, so this high density is clearly not attributable only to our new
candidates. We do note that 5 out of the 7 previously known QSO 2 come from
the paper by Szokoly et al. (\cite{Szokoly04}), which appeared only recently
and could not be accounted for by Gandhi et al (\cite{Gandhi04}). For the same
flux limit and $L_{2 - 10} > 3 \times 10^{44}$ erg/s these authors predict a
surface density $\sim 3$ deg$^{-2}$, while we find 5 sources (3 of them
already known), which implies a density $49^{+33}_{-21}$ deg$^{-2}$. The
decrease in space density at high luminosity is therefore much less than
predicted by their model, namely only a factor $\sim 2$ instead of $\sim 6$.

At even fainter fluxes, Gandhi et al. (\cite{Gandhi04}) quote a value from
their model of 35 deg$^{-2}$ above $f_{\rm} (0.5 - 7 keV) \sim 2 \times
10^{-15}$ erg cm$^{-2}$ s$^{-1}$, while we find a density $\sim 230$
deg$^{-2}$, that is a factor $\sim 6$ larger than their prediction.
 
The population synthesis model of Gilli, Salvati \& Hasinger
(\cite{Gilli01} and private communication) predicts 16 obscured QSO with
intrinsic luminosity $L_{2 - 10} > 3 \times 10^{44}$ erg/s, $N_{\rm H} >
10^{22}$ cm$^{-2}$, and $z > 3$ in the 1 Ms CDF-S. We find 2 known plus 7
candidate sources above these redshift and power limits with $HR \ge
-0.2$. According to Fig. 12 of Szokoly et al. (\cite{Szokoly04}) at $z \ga
3$ an $N_{\rm H} = 10^{22}$ cm$^{-2}$ corresponds to $HR \sim -0.6$ for an
intrinsic $\alpha_{\rm x} = 1$, which means that our definition is more
restrictive. Considering also that the southern GOODS ACS area is $\sim
60\%$ smaller than the CDF-S area, there are strong indications that our
number might be a very solid lower limit to such sources and that therefore
our findings are not inconsistent with the numbers predicted by Gilli et
al. (\cite{Gilli01}).

As mentioned in Sect. \ref{results}, the number of our QSO 2 candidates
depends on the $\log L_{2 - 10} - \log f(2 - 10 keV)/f(R)$ correlation of
Fiore et al. (\cite{Fiore03}), which has an r.m.s. of $\sim 0.5$ dex in X-ray
power. However, even if we consider the worst case scenario in which {\it all}
our estimated X-ray powers are too large by 0.5 dex the number of QSO 2
candidates for the four flux limits in Tab. \ref{counts} are 13, 8, 2, and 1
respectively. In other words, our total densities decrease, at worst, by $\sim
25 - 35\%$ at the lowest fluxes, and are basically unchanged at larger fluxes.


The resolved fraction of the X-ray background due to QSO 2 down to
$f_{\rm} (2 - 8 keV) = 10^{-15}$ erg cm$^{-2}$ s$^{-1}$ is estimated
to be $10^{+2}_{-2}\%$. We have used here the value of the total X-ray
background measured by UHURU and HEAO-1 (Marshall et al. 
\cite{Marshall80}) and the uncertainties reflect the r.m.s. of
the Fiore et al. (\cite{Fiore03}) correlation (the Poissonian error is
even smaller). Note that given the relatively small area of the GOODS
fields we are missing the bright end of the number counts. In fact,
only four of our sources have $f_{\rm} (2 - 8 keV) > 10^{-14}$ erg
cm$^{-2}$ s$^{-1}$ and these reach only $2.6 \times 10^{-14}$ erg
cm$^{-2}$ s$^{-1}$. This fact, the discussion above and the points
detailed in the next section, all mean that our value has to be
regarded as a robust lower limit.

\subsection{Caveats and comments}

In this paper we have employed a statistical method to identify very faint
type 2 AGN. We had also to rely on an empirical technique to estimate the
X-ray powers, which were needed to make sure that the sources we are dealing
with have AGN-like outputs. As such, we can only provide a list of candidates
and not firm classifications. This was obviously expected, as the great
majority of our candidates are so faint that even the largest telescopes
presently available would require extremely long exposures to secure a decent
spectrum. 

However, we believe that our method is robust, as shown by the various checks
we have performed. Importantly, we have been very conservative in our
estimates of the number of type 2 AGN candidates and the surface densities we
have estimated need to be considered lower limits (modulo what discussed in
the previous paragraph), for the following four reasons: 1. our selection
criterion for absorbed sources ($HR \ge -0.2$) will mistakenly discard some
high-z type 2 sources (Sect. \ref{abs}); 2. the same criterion is also more
restrictive than the commonly used one based on $N_{\rm H}$ (Sect.
\ref{type2}); 3. the X-ray powers for previously known sources have not been
corrected for absorption, which means more sources could be above the QSO 2
limit (Sect. \ref{estimate}); and finally, 4. sky coverage effects, once
properly taken into account, will reduce the available area at smaller X-ray
fluxes and therefore increase the source surface density (Sect. \ref{dens}).

\subsection{Next steps enabled with VO tools}

As noted above, a remaining source of uncertainty with the newly discovered
sample is the reliance on an empirical technique in the determination of the
X-ray power of the objects.

In Sect. \ref{testlx} we noted how the 4 colour ACS photometric data have been
used to estimate redshifts for a number of objects in a restricted range.
Future work will enable the use of the 4 band ACS plus IR (VLT/ISAAC and
Spitzer) photometry to determine photometric redshifts over the full GOODS
field. VO technologies are to be employed to facilitate this work. Upgrades to
the AVO capability will include the provision of a ``redshift-determination''
service, which will automate the generation of point spread function matched
multi-band input photometric catalogues via the use of AstroGrid's (Walton,
Lawrence \& Linde \cite{Walton03}) ACE (Astronomical Catalogue Extractor)
service [a Web service providing access to the SExtractor application (Bertin
\& Arnouts \cite{Bertin96})], and feed these into a range of photometric
redshift determination techniques [e.g., Bayesian prior method, Bpz (Ben\'\i
tez \cite{Benitez00}), the neural network technique, ANNz (Collister \& Lahav
\cite{Collister04}) and Hyperz (Bolzonella, Miralles \& Pello
\cite{Bolzonella00})].

It is necessary to include the IR data in order to obtain
reliable photometric redshifts, and with the
additional photometric bands, the effects of possible dust extinction
will also be identifiable (Rowan-Robinson \cite{Rowan03}).

\section{Conclusions}

We have used Virtual Observatory tools to identify obscured AGN much fainter
than previously known ones, going beyond what is currently possible using the
``classical'' approach of classifying sources by taking spectra of them even
using the largest telescopes available today. The fact that we have obtained
scientifically useful and cutting-edge results is proof that VO tools have
evolved beyond the demonstration level to become respectable research
tools. The VO is already enabling astronomers to reach into new areas of
parameter space with relatively little effort.

Our main results can be summarized as follows:

\begin{enumerate}

\item By employing publicly available, high-quality X-ray and optical data we
have discovered 68 type 2 AGN candidates in the two GOODS fields. Their X-ray
powers have been estimated by using a previously known correlation between
X-ray luminosity and X-ray-to-optical flux ratio. Thirty-one of our candidates
have high luminosity ($L_{\rm 2 - 10 keV} > 10^{44}$ erg/s) and therefore
qualify as QSO 2, that is optically obscured quasars. Based on their derived
X-ray powers, our candidates are likely to be at relatively high redshifts, $z
\sim 3$, with the QSO 2 at $z \sim 4$.

\item We have tested our method and results extensively and find them to be
very robust. In particular: a) our approach recovers most ($\sim 97\%$) of the
previously known type 2 AGN in the GOODS fields; b) the X-ray power estimates
agree very well with those derived from spectroscopic redshifts for
the non-type 1 AGN in the GOODS fields; c) the redshifts derived from our
estimated powers are consistent with those measured for previously known
non-type 1 AGN; d) the spectrum of our brightest northern candidate, recently
made public as part of the Team Keck Treasury Redshift Survey, confirms our
type 2 classification as it is basically featureless, with a measured redshift
extremely close to our estimate.

\item By going $\sim 3$ magnitudes fainter than previously known type 2 AGN we
are sampling a region of redshift -- power space so far unreachable with
classical methods. Our method brings to 40 the number of QSO 2 in the
GOODS fields, an improvement of a factor $\sim 4$ when compared to the 9 such
sources previously known. The relatively large QSO 2 number density we derive
($\ga 330$ deg$^{-2}$ for $f(0.5 - 8 keV) \ge 10^{-15}$ erg cm$^{-2}$
s$^{-1}$, $\sim 30\%$ of which is made up of already known sources) suggests
that its value at faint flux limits has been underestimated. 

\end{enumerate}

For the handful of our candidates with $R \la 26$ spectroscopy with a
class 8 - 10 m telescope is still within reach, albeit with relatively
long ($\la 10$ h) exposure times. We plan to follow these up and
confirm (or refute) their classification in the near future.

In closing we note that this paper represents the first significant published
science result that has been fully enabled via end-to-end use of VO tools and
systems.

VO initiatives are in their early stages. Significant progress is being made
both by the AVO and other national VO projects. Construction of a truly
pervasive system is beginning to provide the end user access to a powerful
combination of data access, applications embedded within user workflows,
running over high speed networks on powerful compute resources. Early use,
exploiting a prototype system in the USA, constructed by the US-VO
project\footnote{\tt http://www.us-vo.org} to simplify the cross matching of
Two Micron All Sky Survey and Sloan Digital Sky Survey source catalogues has
enabled Berriman et al. (\cite{Berriman03}) to discover a small number of
hitherto undiscovered Brown Dwarf candidates. Their work demonstrated
possibilities, but did not alter the scientific understanding of that
particular problem.

In this paper we have shown how, even with the AVO's early prototype, enough
capability and access to data is available to make the scientific mining of
the, in this case GOODS, data significantly easier for the ``general''
astronomer. This is because gains in interoperability simplify the acquisition
of the data, and enable quick interoperation of a number of common, but
computationally intensive, tasks such as cross matching and multi layer
visualisation of both image and catalogue data.

With the rapid evolution of the VO, science discovery will be routinely
performed using VO techniques, an early example of which is described here.

The AVO prototype used in this paper is freely available for
download\footnote{\tt
http://www.euro-vo.org/twiki/bin/view/Avo/SwgDownload}. 

\begin{acknowledgements}

We thank Fabrizio Fiore for providing the data on which the X-ray
power -- X-ray-to-optical flux ratio correlation was based, Roberto
Gilli, Vincenzo Mainieri, and Paolo Tozzi for useful
discussions, Sebastien Derriere for assistance with the false match
rate estimates, and the AVO team for their superb work, without which
this paper would have not been possible. Finally, a sincere ``thank
you'' to the many people who have produced the data on which this
paper is based, particularly the GOODS, CDF-S, and Penn State
Teams. This research has made use of the CDS Vizier catalogue tool,
SIMBAD and Aladin sky atlas services.

\end{acknowledgements}

\appendix
\section{False Match Estimate}
\label{app1}

We estimated the number of false matches using the formalism described 
in Derriere (\cite{Derriere01}). The expected total number of matches $D$, 
for $N_1$ objects with positional uncertainty $\sigma$ against a uniformly 
distributed set of objects with density $\lambda$ is:

\begin{eqnarray}
D(d,N_1,\lambda,\sigma,N)= \nonumber \\
N_1\Phi(d,\lambda)+N\Phi(d,\lambda)\left[\frac{1+2\pi\lambda\sigma^2}{2\pi
\lambda\sigma^2}\exp\left(-\frac{d^2}{2\sigma^2}\right)-1\right]
\end{eqnarray}

where $d$ is the distance between matches, $\Phi(d,\lambda)$ is the 2-d
Poisson density function\footnote{$\Phi(d,\lambda)=2\pi\lambda
d\exp(-\pi\lambda d^2)$} and $N$ is the number of objects with a true
counterpart. $D$ may also be expressed a sum of three components,
$\alpha$\footnote{$\alpha=\frac{N}{2\pi\lambda\sigma^2}\Phi(d,\lambda)\exp\left(
-\frac{d^2}{2\sigma^2}\right)$} the number of true matches,
$\beta$\footnote{$\beta=N\Phi(d,\lambda)\exp\left(-\frac{d^2}{2\sigma^2}\right)$
} the objects with a true partner but not incorrectly assigned and
$\psi$\footnote{$\psi=(N_1-N)\Phi(d,\lambda)$} the number of objects that do
not have a counterpart but have been matched.

\begin{figure}
   \centering
   \resizebox{\hsize}{!}{\includegraphics{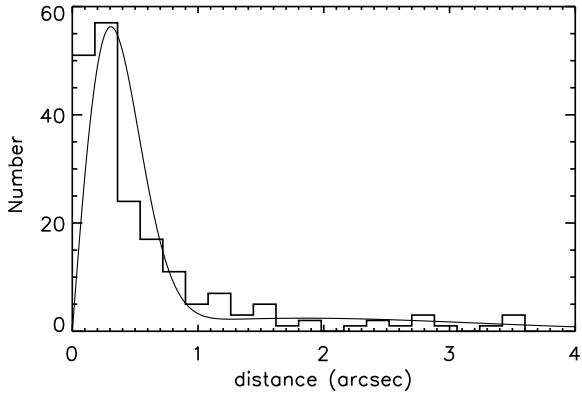}}
      \caption{The histogram of distances between closest match optical and X-ray 
      sources overlaid with the best fitting model}
         \label{d_hist}
   \end{figure}

Using our known values of $N_1$: the number of absorbed sources (203),
$\lambda$: the optical source density of the GOODS catalogues
(0.0469~arcsec$^{-2}$), and $\sigma$: the median positional error of the X-ray
positions of the absorbed sources (0.31\arcsec) we vary $N$ to perform a
maximum likelihood fit of the expected distribution of matches $D$ to the
observed histogram of distances between closest match optical and X-ray
sources out to the initial 3.5\arcsec\ threshold radius. The best fit occurs
for $N=160$. Fig. \ref{d_hist} shows the histogram of distances between
closest match optical and X-ray sources overlaid with the best fitting model.

Integrating $D$ and $\alpha$ out to 1.25\arcsec\ we find that 92\% of the
matches out to this radius are expected to be true according to the model,
making our false match fraction 8\%. Note that this approach neglects
clustering, which would slightly increase this value.

We also estimated the number of false matches by comparing the number of X-ray
to optical matches we find using the correct coordinates, to the number of
matches we find when a shift (of 20\arcsec\, much larger than the typical
match radius and the spacing between optical sources) is applied to the X-ray
coordinates. Within 1.25\arcsec\, we find 188 matches and 38 matches when the
coordinates are shifted. Our selection of real matches however, also included
the distance/error criterion which means that a number of matching sources
within 1.25\arcsec\ of the X-ray source were discarded. This criterion should
also be taken into consideration when estimating the rate of false matches. We
estimate this effect on the false match rate by calculating the rate of
optical sources within 1.25\arcsec\ of the real X-ray position which have
distance/error $>1$ (10), and then subtract this from the rate of false matches we
find in the offset cross-match. Therefore the false match fraction using this
method should be (38-10)/188 = 15\%.

The fraction of false matches should then be in the range $8 - 15\%$.

\end{document}